\documentclass[aps,amsmath,amssymb,prl,twocolumn,superscriptaddress,unsortedaddress]{revtex4}
\usepackage{amsmath}
\usepackage{amssymb}
\usepackage{float}
\usepackage{dcolumn}
\usepackage{bm}
\usepackage{epsf}
\usepackage{graphicx}
\usepackage{amsfonts}
\usepackage{sidecap}
\sidecaptionvpos{figure}{c}

\newcommand{\etal}{{\it et al.}}
\begin{document}

\title{Gapless~surface~Dirac~cone~in antiferromagnetic~topological~insulator~MnBi$_2$Te$_4$}

\author{Yu-Jie~Hao}
\affiliation {Shenzhen Institute for Quantum Science and Engineering (SIQSE) and Department of Physics, Southern University of Science and Technology (SUSTech), Shenzhen, Guangdong 518055, China}

\author{Pengfei~Liu}
\affiliation {Shenzhen Institute for Quantum Science and Engineering (SIQSE) and Department of Physics, Southern University of Science and Technology (SUSTech), Shenzhen, Guangdong 518055, China}

\author{Yue~Feng}
\affiliation {Shenzhen Institute for Quantum Science and Engineering (SIQSE) and Department of Physics, Southern University of Science and Technology (SUSTech), Shenzhen, Guangdong 518055, China}

\author{Xiao-Ming~Ma}
\affiliation {Shenzhen Institute for Quantum Science and Engineering (SIQSE) and Department of Physics, Southern University of Science and Technology (SUSTech), Shenzhen, Guangdong 518055, China}

\author{Eike~F.~Schwier}
\affiliation {Hiroshima Synchrotron Radiation Center, Hiroshima University, 2-313 Kagamiyama, Higashi-Hiroshima 739-0046, Japan}

\author{Masashi~Arita}
\affiliation {Hiroshima Synchrotron Radiation Center, Hiroshima University, 2-313 Kagamiyama, Higashi-Hiroshima 739-0046, Japan}

\author{Shiv~Kumar}
\affiliation {Hiroshima Synchrotron Radiation Center, Hiroshima University, 2-313 Kagamiyama, Higashi-Hiroshima 739-0046, Japan}

\author{Chaowei~Hu}
\affiliation {Department of Physics and Astronomy and California NanoSystems Institute, University of California, Los Angeles, California 90095, USA}

\author{Rui'e~Lu}
\affiliation {Shenzhen Institute for Quantum Science and Engineering (SIQSE) and Department of Physics, Southern University of Science and Technology (SUSTech), Shenzhen, Guangdong 518055, China}

\author{Meng~Zeng}
\affiliation {Shenzhen Institute for Quantum Science and Engineering (SIQSE) and Department of Physics, Southern University of Science and Technology (SUSTech), Shenzhen, Guangdong 518055, China}

\author{Yuan~Wang}
\affiliation {Shenzhen Institute for Quantum Science and Engineering (SIQSE) and Department of Physics, Southern University of Science and Technology (SUSTech), Shenzhen, Guangdong 518055, China}

\author{Zhanyang~Hao}
\affiliation {Shenzhen Institute for Quantum Science and Engineering (SIQSE) and Department of Physics, Southern University of Science and Technology (SUSTech), Shenzhen, Guangdong 518055, China}

\author{Hong-Yi~Sun}
\affiliation {Shenzhen Institute for Quantum Science and Engineering (SIQSE) and Department of Physics, Southern University of Science and Technology (SUSTech), Shenzhen, Guangdong 518055, China}

\author{Ke~Zhang}
\affiliation {Hiroshima Synchrotron Radiation Center, Hiroshima University, 2-313 Kagamiyama, Higashi-Hiroshima 739-0046, Japan}

\author{Jiawei~Mei}
\affiliation {Shenzhen Institute for Quantum Science and Engineering (SIQSE) and Department of Physics, Southern University of Science and Technology (SUSTech), Shenzhen, Guangdong 518055, China}

\author{Ni~Ni}
\affiliation {Department of Physics and Astronomy and California NanoSystems Institute, University of California, Los Angeles, California 90095, USA}

\author{Liusuo~Wu}
\affiliation {Shenzhen Institute for Quantum Science and Engineering (SIQSE) and Department of Physics, Southern University of Science and Technology (SUSTech), Shenzhen, Guangdong 518055, China}

\author{Kenya~Shimada}
\affiliation {Hiroshima Synchrotron Radiation Center, Hiroshima University, 2-313 Kagamiyama, Higashi-Hiroshima 739-0046, Japan}

\author{Chaoyu~Chen}
\email{chency@sustech.edu.cn}
\affiliation {Shenzhen Institute for Quantum Science and Engineering (SIQSE) and Department of Physics, Southern University of Science and Technology (SUSTech), Shenzhen, Guangdong 518055, China}

\author{Qihang~Liu}
\email{liuqh@sustech.edu.cn}
\affiliation {Shenzhen Institute for Quantum Science and Engineering (SIQSE) and Department of Physics, Southern University of Science and Technology (SUSTech), Shenzhen, Guangdong 518055, China}
\affiliation {Guangdong Provincial Key Laboratory for Computational Science and Material Design, Southern University of Science and Technology, Shenzhen, Guangdong 518055, China}

\author{Chang~Liu}
\email{liuc@sustech.edu.cn}
\affiliation {Shenzhen Institute for Quantum Science and Engineering (SIQSE) and Department of Physics, Southern University of Science and Technology (SUSTech), Shenzhen, Guangdong 518055, China}

\date{\today}

\clearpage

\begin{abstract}

The recent discovered antiferromagnetic topological insulators in Mn-Bi-Te family with intrinsic magnetic ordering have rapidly drawn broad interest since its cleaved surface state is believed to be gapped, hosting the unprecedented axion states with half-integer quantum Hall effect. Here, however, we show unambiguously by using high-resolution angle resolved photoemission spectroscopy that a gapless Dirac cone at the (0001) surface of MnBi$_2$Te$_4$ exists inside the bulk band gap. Such unexpected surface state remains unchanged across the bulk N\'{e}el temperature, and is even robust against severe surface degradation, indicating additional topological protection. Through symmetry analysis and \textit{ab-initio} calculations we consider different types of surface reconstruction of the magnetic moments as possible origins giving rise to such linear dispersion. Our results unveil the experimental topological properties of MnBi$_2$Te$_4$, revealing that the intrinsic magnetic topological insulator hosts a rich platform to realize various topological phases by tuning the magnetic/structural configurations, and thus pushed forward the comprehensive understanding of magnetic topological materials.

\end{abstract}

\maketitle

\subsection{I. Introduction}

The integration of magnetic order and topological nontriviality catches wide attention since the dawn of the topological era in condensed matter physics \cite{Yulin_Science, Suyang_NP, Tokura}. In these systems, the absence of time-reversal symmetry ($\mathcal{T}$) brings about a series of exotic quantum phases such as Chern insulator \cite{Regnault} and axion insulator \cite{Zhang_field_theory}, leading to potential applications in the fields of spintronics and quantum computing. As a renowned example, the quantum anomalous Hall (QAH) effect in Chern insulators promises novel emergent phenomena such as Majorana fermions and anyon statistics \cite{YuRui}. Another distinct topological phase is the axion insulator state, signified by a gapped surface state by magnetization but half-quantized surface Hall conductance, which was proposed to host the topological magnetoelectric effect and axion electrodynamics \cite{Zhang_field_theory, FangChen}. QAH insulator was first discovered in a magnetically doped topological insulator (TI) at an ultralow temperature of 30 mK \cite{Xue_QAHE}. Proposals of heterostructure engineering based on magnetic insulators and TIs are also expected to realize the QAH effect through magnetic proximity effects, but are challenging in the material choice and interface fabrication \cite{Zhang_field_theory}. On the other hand, the realization of an axion insulator requires a TI layer sandwiched by two magnetic layers whose moments point to opposite directions \cite{Xiao}. Though the QAH effect and the axion state have been discovered in TIs and magnetic topological heterostructures where ferromagnetically ordered moments are induced by chemical doping \cite{Xue_QAHE, Xiao, Checkelsky, Kou, ChangCuizu}, an intrinsic, stoichiometric magnetic topological insulator (MTI) is highly desired in both cases, as the emerging temperatures of these macroscopic quantum states would otherwise be severely suppressed due to disorder.

\begin{SCfigure*}
\centering
\includegraphics[width=13cm]{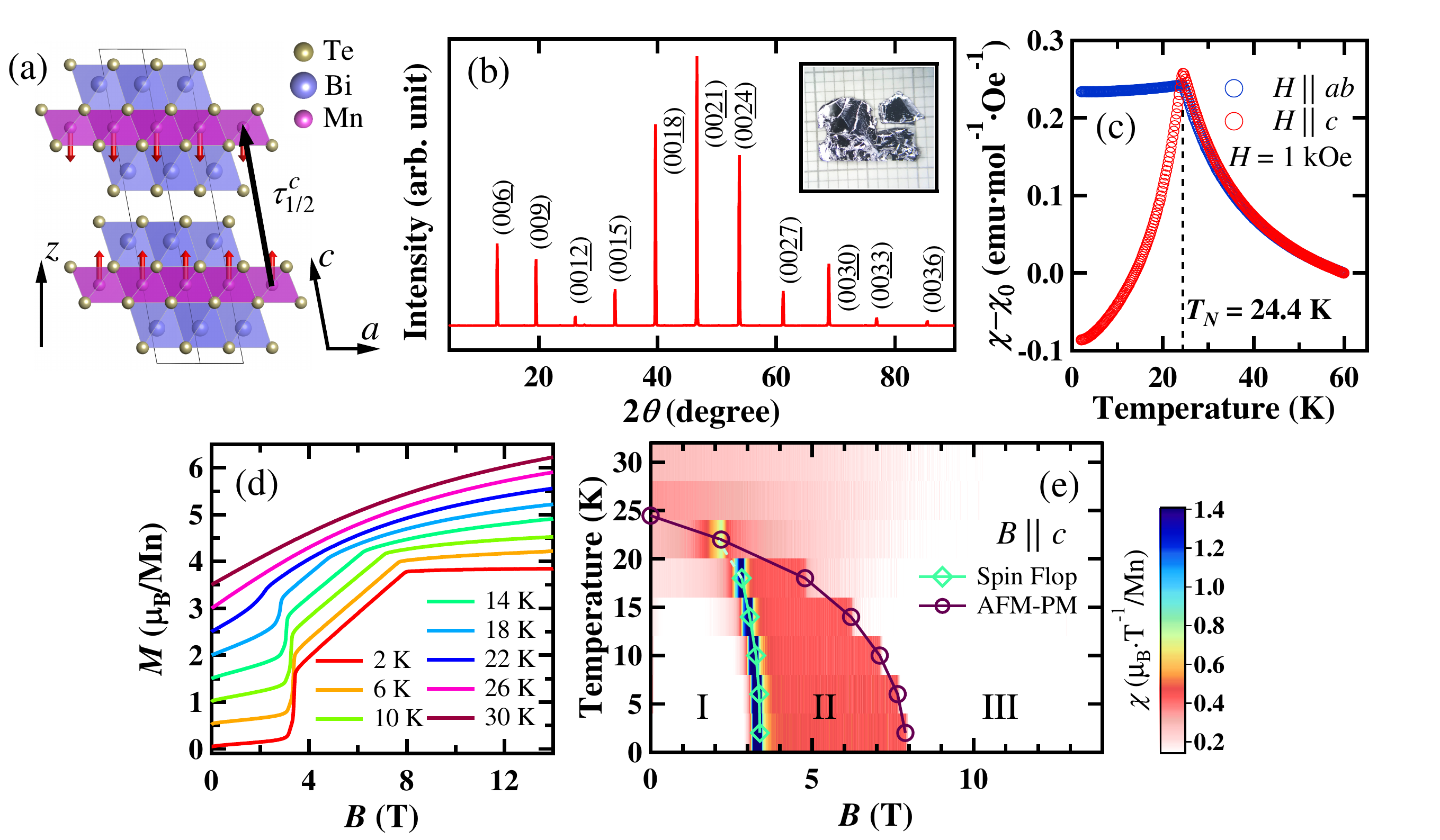}
\caption{Crystal characterization and magnetic responses of MnBi$_2$Te$_4$. (a) Crystal structure and A-type AFM magnetic configuration. (b) Single crystal x-ray diffraction data. Inset: crystal against an mm grid. (c) Magnetization curves in two different configurations, $H \| ab$ and $H \| c$. The susceptibility at 60 K ($\chi_0$) is subtracted. (d) Field dependence of magnetization \textit{M}, measured at different \textit{T}, with field along the \textit{c} axis. (e) The field-temperature phase diagram with applied field along the \textit{c} axis. As \textit{B} increases, the system evolves from an A-type AFM state (I) to a spin-flop AFM state (II), then to a high-\textit{T} PM state (III).}\label{Fig1}
\end{SCfigure*}

For both of QAH insulators and axion insulators, materials that best fit the above requirement should be magnetic layered compounds having net magnetization and zero magnetization, respectively. Interestingly, both conditions can be achieved in a single material base comprising A-type antiferromagnetism (AFM) with out-of-plane magnetic moments, in which the two distinct topological phases can be switched simply by controlling the number of layers. Such a three-dimensional (3D) material base is a AFM TI characterized by a novel $\mathbb{Z}_2$ topological invariant protected by the product of $\mathcal{T}$ and a half-cell translation along the \textit{c} axis $\tau_{1/2}^c$, named $\mathcal{S}^c = \mathcal{T}\tau_{1/2}^c$ \cite{Mong}. While layered AFM insulators like CrI$_3$ fulfills the magnetic structure but lacks a nonzero topological invariant \cite{Huang}, a novel family of Van der Waals layered single crystalline materials MnBi$_{2n}$Te$_{3n+1}$ ($n = 1, 2, ...$) \cite{Aliev, LiuQihang}, exemplified by MnBi$_2$Te$_4$ \cite{LeeSH, Otrokov, Zeugner, Gong_CPL, LiJH, Otrokov2, YanJiaqiang, Deng, YayuWang, GeJ, Vidal, YanJiaqiang2, HuCW, ChenBo, Cui, ZhangDQ, LiJH2, LeeDS, WuJZ}, is established very recently to be stoichiometric TIs with an A-type AFM ground state. One basic magnetic building block of MnBi$_2$Te$_4$ consists of seven atomic layers that arranges as Te-Bi-Te-Mn-Te-Bi-Te \cite{LeeSH, Otrokov, Zeugner, Gong_CPL}, named a septuple layer (SL) [Fig. \ref{Fig1}(a)]. Its magnetic moments in the bulk are theoretically predicted \cite{LiJH, Otrokov2}, and then confirmed by neutron diffraction experiments \cite{YanJiaqiang}, to be ferromagnetically (FM) ordered within a Mn plane pointing along the out-of-plane \textit{z}-direction but antiferromagnetically aligned between adjacent Mn layers [Fig. \ref{Fig1}(a)]. It was experimentally found that few-SL films of MnBi$_2$Te$_4$ can realize not only the axion state but a topological transition between the axion state and the QAH state, when switching the number of SLs between even and odd numbers \cite{Deng, YayuWang}. Furthermore, nine layers of MnBi$_2$Te$_4$ is experimentally reported to be a higher-order Chern insulator \cite{GeJ}. Such rich emergent physics render MnBi$_2$Te$_4$ an ideal platform for studying the interplay between magnetism and topology.

Now that several theoretical predictions and experimental observations point to a simple A-type AFM ground state in bulk MnBi$_2$Te$_4$, there are still substantial discrepancies between the ideal scenario and realistic quantum transport behaviors. For example, while the QAH effect was predicted to appear in odd number of MnBi$_2$Te$_4$ layers with uncompensated A-type AFM configuration, such effect was observed experimentally only under a strong magnetic field ($>5$ T) that forces the AFM ground state to a FM one \cite{Deng, YayuWang}. This implies that the inherent magnetic configuration, including the magnitude, orientation, domain, and bulk-surface correspondence, adds complication to the full understanding of intrinsic MTIs such as MnBi$_2$Te$_4$. Since the key property for the realization of axion state is the gapped Dirac cone induced by intrinsic magnetization, it is crucial to verify the existence of such a gapped surface state in bulk MnBi$_2$Te$_4$. In this paper, we profile the topological nature of MnBi$_2$Te$_4$ by our experimental discovery of the unexpected bulk-surface correspondence using high-resolution angle resolved photoemission spectroscopy (ARPES). Unlike previous theoretical predictions and experimental observations claiming a sizable magnetic gap at the (0001) surface state where $\mathcal{S}^c$ is broken \cite{LeeSH, Otrokov, Zeugner, Vidal}, we show unambiguously that there exist an X-shaped, gapless, Dirac cone at this surface, traversing the bulk band gap of MnBi$_2$Te$_4$. This state is intrinsic to the MnBi$_2$Te$_4$ crystals, reproducible in all samples we measured, free of $k_z$ dispersion, unchanged across the bulk magnetic ordering temperature, and is even robust against severe surface degradation. The gapped bands observed by previous works near the Dirac point, on the other hand, are proven to be of bulk nature, having clear $k_z$ dispersion. By performing symmetry analysis and density functional theory (DFT) calculations, we attribute the origin of the observed gapless Dirac cone to surface-mediated reconstruction of magnetic moments that differ from the bulk, with the discussion of several proposed occasions including spin disorder, A-type AFM with in-plane moment and intralayer AFM. The possibility of surface structural deformation is also discussed. Our work reveals an important factor that can significantly affect the topological property of MnBi$_2$Te$_4$, i.e., the surface magnetic/structural reconstruction, and thus brings about a more comprehensive understanding of magnetic topological materials in general.

\begin{SCfigure*}
\centering
\includegraphics[width=12cm]{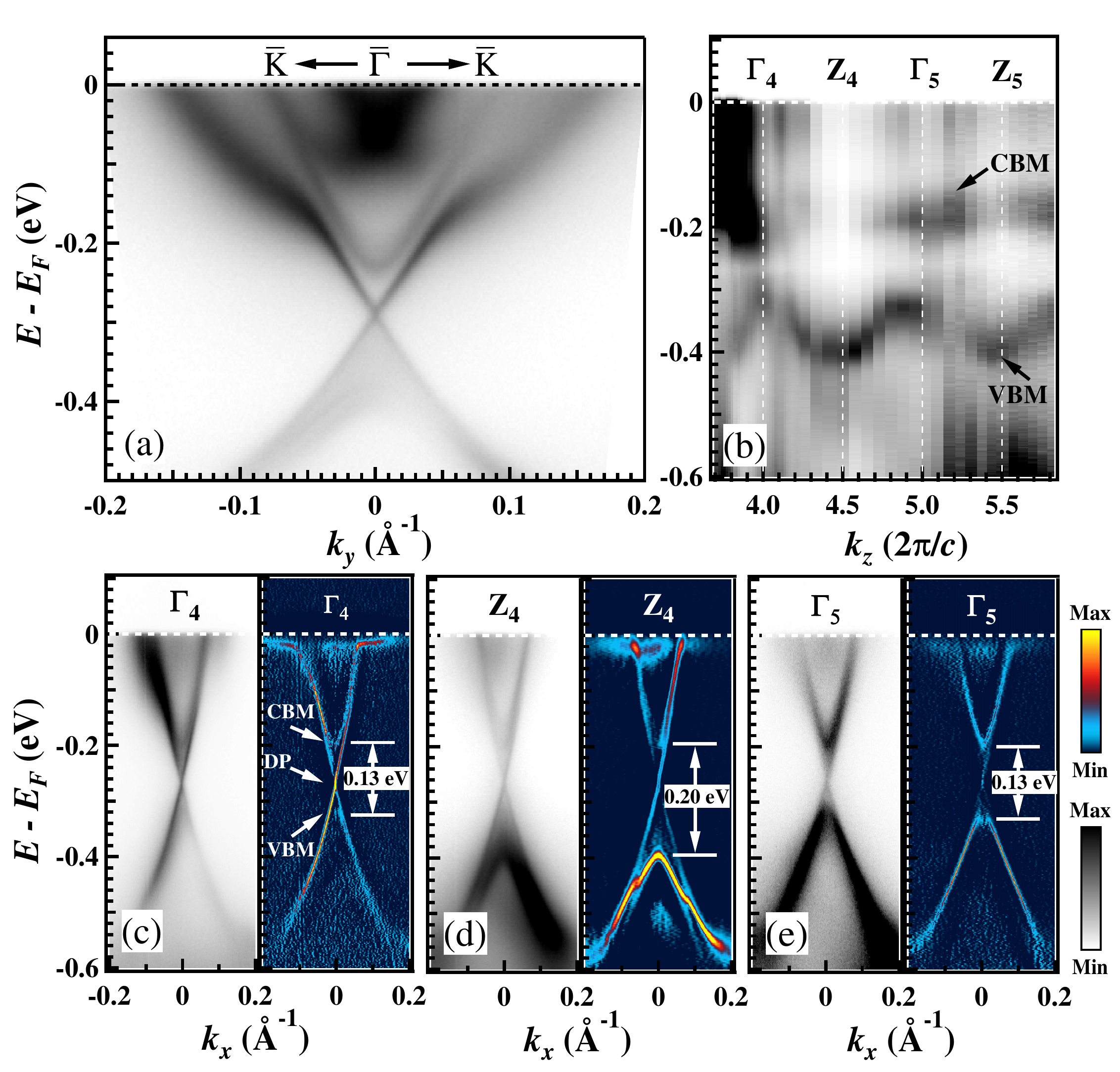}
\caption{Surface and bulk electronic structure of MnBi$_2$Te$_4$. (a) A typical ARPES $k$-$E$ map along $\bar{K}$-$\bar{\Gamma}$-$\bar{K}$ ($k_y$), taken at 10 K under photon energy $h\nu = 6.3$ eV. A linear, X-shaped, gapless state exists between the valence and the conduction bands. The Dirac point energy locates at $E_D = 288$ meV for this sample. (b) $k_z$ dispersion map at $\bar{\Gamma}$, taken with 6 to 20 eV photons. VBM: valence band maxima; CBM: conduction band minima. Periodic dispersion pattern on the VBM is seen clearly. The bulk $\Gamma$ and $Z$ points are determined by assigning an inner potential $V_0 = 9$ eV, estimated from the total bandwidth of the valence bands. (c)-(e) $k$-$E$ maps along $\bar{M}$-$\bar{\Gamma}$-$\bar{M}$ ($k_x$) taken at the $\Gamma_4$, $Z_4$ and $\Gamma_5$ points marked in (b) (correspond to $h\nu = 7.5$, 10.5 and 13.5 eV, respectively). It is clear that the gapless state forming the Dirac cone has no $k_z$ dispersion, while the VBM evolves from -0.33 eV ($\Gamma$) to -0.4 eV ($Z$), consequently changing the bulk gap from 0.13 to 0.20 eV.}\label{Fig2}
\end{SCfigure*}

\subsection{II. Crystal and magnetic properties of bulk MnBi$_2$Te$_4$}

We begin our discussion by presenting the physical properties measured on the MnBi$_2$Te$_4$ samples used in our ARPES measurements. It is important to point out that our samples were grown by two different research groups (UCLA and SUSTech) using slightly different growth procedures, yet the transport, magnetic, and ARPES measurements reveal quantitatively the same results, each on multiple samples, signifying the reliability and repeatability of our data. Fig. \ref{Fig1}(b) presents the single crystal x-ray diffraction data, as well as the appearance of the crystals, which agree quantitatively with those in the literature. No signal from other crystalline phases is seen. This proves that our photoemission data does not come from impurity phases.

\begin{figure*}
\centering
\includegraphics[width=13cm]{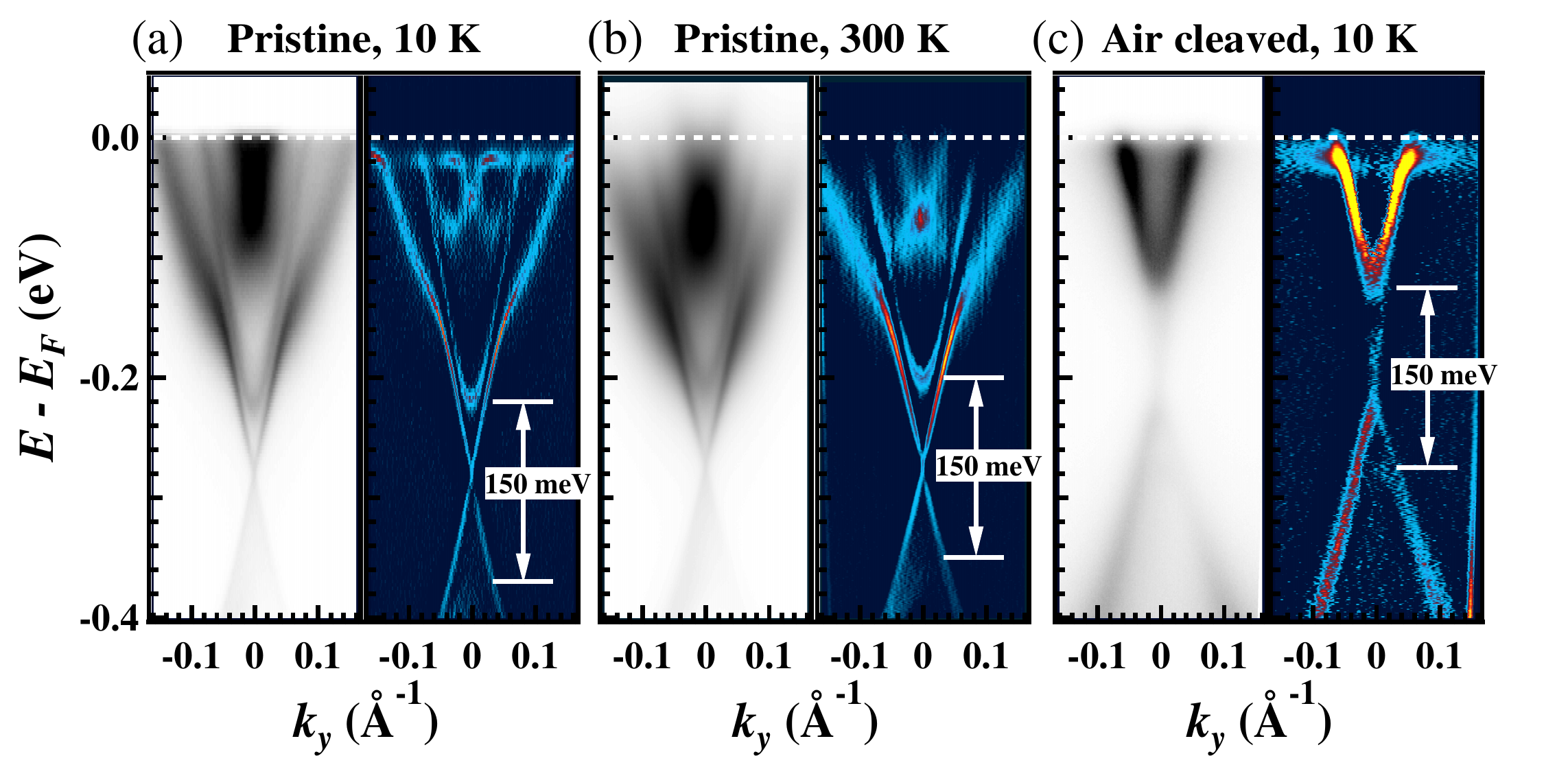}
\caption{Robustness of the Dirac surface state. The figure shows ARPES raw (left) and second derivative (right) $k$-$E$ maps taken with 6.3 eV laser light along $\bar{K}$-$\bar{\Gamma}$-$\bar{K}$ ($k_y$) for (a) a pristine sample cleaved and measured at 10 K (below $T_N$), (b) a pristine sample cleaved and measured at 300 K (way above $T_N$), and (c) a sample cleaved in air at room temperature, and measured at 10 K. Despite the overall carrier doping induced by different cleaving conditions, the gapless Dirac cone is clearly seen for all cases, along with an unchanged bulk band gap sized 150 meV.}\label{Fig3}
\end{figure*}

In Figs. \ref{Fig1}(c)-(e) we present the magnetic measurement results, demonstrating that MnBi$_2$Te$_4$ has an AFM ground state and a rich magnetic phase diagram. Fig. \ref{Fig1}(c) shows the magnetization versus temperature curves for two different configurations, $H \| ab$ and $H \| c$. An AFM-paramagnetic (PM) transition is found at $T_N = 24.4$ K, consistent with data from other groups. Fig. \ref{Fig1}(d) illustrates the isothermal magnetizations of MnBi$_2$Te$_4$ as a function of applied field along the \textit{c} axis ranging from 2 to 30 K. All the magnetization curves for $T < T_N$ show an abrupt change around the field between 2 T and 3.6 T. This suggests the occurrence of a spin-flop type transition, below which the spin direction of Mn ions turn perpendicular to the easy axis (\textit{c} axis). Finally, the magnetization approaches to saturation around $M = 3.8$ $\mu_B$/Mn at 2 K above 8 T, well consistent with previous results \cite{YanJiaqiang, YanJiaqiang2, HuCW}. The field-temperature phase diagram of MnBi$_2$Te$_4$ is depicted in Fig. \ref{Fig1}(e). Below $T_N$, the critical field of spin-flop transition $B_{c1}$ divides the phase diagram into two regions. At region I, MnBi$_2$Te$_4$ shows an A-type AFM order consisting of FM layers coupled antiferromagnetically along the \textit{c} axis \cite{YanJiaqiang}. Above $B_{c1}$ (region II), it is possible that the moments first turn perpendicular to the \textit{c} axis due to the lower ground energy, then rotates continuously towards the field direction. When the critical field $B_{c2}$ is reached, all the moments are polarized along the applied field (region III).

\subsection{III. Robust surface Dirac cone by ARPES measurements}

We show in Fig. \ref{Fig2} the electronic structure of MnBi$_2$Te$_4$ obtained by high resolution ARPES experiments. Fig. \ref{Fig2}(a) illustrates a typical ARPES $k$-$E$ cut through the surface zone center $\bar{\Gamma}$ we obtained with high resolution laser ARPES \cite{Iwasawa} below $T_N$ ($h\nu = 6.3$ eV, $T = 10$ K). Even at the first glance, one notices that there exists undoubtedly a gapless state between the electronlike and holelike conduction and valence bands, whose dispersion is even more linear than conventional TIs like Bi$_2$Se$_3$ and Bi$_2$Te$_3$. The two branches of this state intersect at $\bar{\Gamma}$ at a binding energy of 0.290 eV, forming a prototypical Dirac cone at $\bar{\Gamma}$ without any trace of gap. This band is one of the sharpest electronic states ever seen in topological materials, with a full-width half-maximum (FWHM) of 0.010 $\textrm{\AA}^{-1}$ (detailed in \cite{Supplement}). Please note that an ungapped Dirac cone is also reported in Ref. \cite{LiJH} for few-SL films of MnBi$_2$Te$_4$, yet the measurements there were done only at the PM state above $T_N$. The main purpose of the present paper is to study the spectroscopic properties of this state, and to propose, based on reasonable symmetry arguments and DFT calculations, the origin of this state.

Having established the existence of the gapless Dirac state, an important question is whether this gapless band remains unchanged for all $k_z$s in the Brillouin zone (i.e., it is a 2D state), or it gradually opens a gap as $k_z$ varies (i.e., it is a 3D state). We prove that this state is in fact a quasi-2D state with little $k_z$ dispersion by performing a detailed photon energy dependent ARPES map from 6 to 20 eV, corresponding to $3.7 < k_z < 5.8$ (in unit of $2\pi/c$), covering more than two Brillouin zones in the $k_z$ direction [Fig. \ref{Fig2}(b)]. Note that the lattice constant $c$ here represents the height of a single SL of MnBi$_2$Te$_4$ ($c = 1.37$ nm), which is about 1/3 the height of the conventional unit cell. For all $k_z$ values within this range where the Dirac state is resolvable, we see negligible change in the ($k, E$) position of the Dirac band \cite{Supplement}. Therefore, the Dirac state behave as a quasi-2D, surfacelike electronic state. On the other hand, the bands that form a gap at $\bar{\Gamma}$, especially the one below the Dirac band, shows clear periodic dispersion along $k_z$ [Fig. \ref{Fig2}(b)]. As a result, the size of the $\bar{\Gamma}$ gap varies between 0.13 eV at the bulk $\Gamma$, and 0.20 eV at $Z$ [Figs. \ref{Fig2}(c)-(e)]. In light of this behavior, we assign this gapped band to be a bulklike electronic state.

Next we study whether this Dirac state remains across the magnetic transition temperature, and whether it is robust against considerable surface perturbation. Fig. \ref{Fig3} gives positive answers to both questions. When an as-grown, pristine MnBi$_2$Te$_4$ crystal is cleaved \textit{in situ} at 10 K (below $T_N$), it shows a clear ungapped Dirac cone whose Dirac point lies at $E_b = 0.28$ eV [Fig. \ref{Fig3}(a)] \cite{Note}. In case of 6.3 eV photons, the bulk gap measures to be 150 meV. When we rise the temperature to 300 K [Fig. \ref{Fig3}(b)], the Dirac point energy changes to $E_b = 0.27$ eV, yet the gapless nature of the Dirac band, as well as the size of the bulk gap, remains. Therefore, being in the AFM ground state or the high-temperature PM state do not seem to affect the integrity of the cone. To further test the robustness of the cone, we cleave a sample in air at room temperature before loading it into the vacuum chamber, and measure the band structure of the disturbed surface at 10 K [Fig. \ref{Fig3}(c)]. Although the band structure becomes significantly $p$-doped compared with the pristine one, the Dirac surface state is still visible, without any sign of gap opening. The bulk gap also keeps its size of 150 meV. The message here is that this Dirac state is as robust as those in prototypical nonmagnetic TIs like Bi$_2$Se$_3$ \cite{Chaoyu}. Therefore, it is highly likely that this state is topologically protected by a symmetry of the crystal.

\subsection{IV. Proposed gapless Dirac cone from different surface magnetic structures}

\begin{figure}
\centering
\includegraphics[width=9cm]{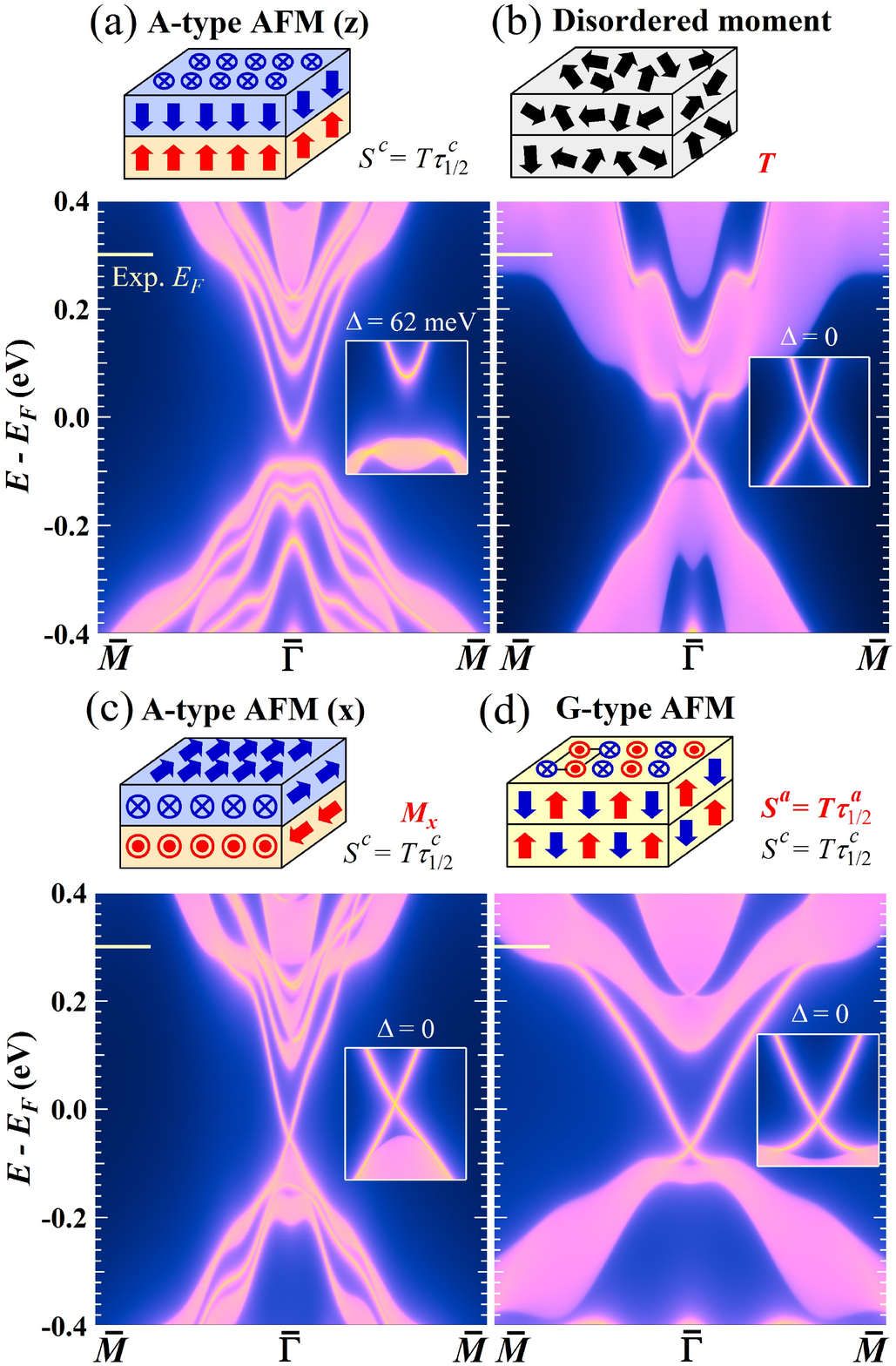}
\caption{DFT-calculated surface states of MnBi$_2$Te$_4$ for four prototypical magnetic configurations, i.e., (a) A-type AFM with the magnetic moments along the $z$ axis, (b) disordered magnetic moments, (c) A-type AFM with the magnetic moments along the $x$ axis, and (d) G-type AFM. The top drawing in each panel specifies the magnetic configuration and the symmetries that preserve in bulk (black) and (0001) surface (red). $\mathcal{T}$: time reversal; $\tau_{1/2}^i$ ($i = a, b, c$): half-cell translation along $i$; $\mathcal{M}$: mirror plane.}\label{Fig4}
\end{figure}

Next we discuss the possible physical origin of the gapless (0001) surface Dirac cone in MnBi$_2$Te$_4$ from the perspective of surface spin reorientation. For 3D magnetic-doped TIs, it is reported that the helical surface electrons can induce a FM order at the surface through RKKY interaction even when the bulk is not magnetically ordered \cite{Wray, LiuQ, ZhuJJ, Rosenberg}. In MnBi$_2$Te$_4$, previous neutron diffraction measurements confirmed an A-type AFM spin configuration with the magnetization along the $c$ axis \cite{YanJiaqiang}, which supports a massive surface Dirac cone if the bulk magnetic configuration remains at the surface [Fig. \ref{Fig4}(a)]. Therefore, our results suggest that the surface-mediated spin configuration at the few top layers differs from that in the bulk state, hosting topology-protected gapless surface states which can be detected by our surface-sensitive ARPES technique. In the following we consider several possible magnetic states of the surface layers that can support the linear-dispersed, gapless surface Dirac cone, and then discuss their chance to happen based on the current experimental observations and our corresponding total energy calculations. They are summarized in Fig. \ref{Fig4}(b)-(d) and Table \ref{Table1}.

Before proposing the possibilities, we make a basic assumption that the gapless surface states are protected by $\mathcal{T}$, crystalline symmetry or their combinations. Starting from nonmagnetic MnBi$_2$Te$_4$ having the same space group of Bi$_2$Te$_3$ ($R\bar{3}m$), there are totally five types of symmetry operations, i.e., 3-fold rotation along the $z$ axis $\mathcal{C}_{3z}$, 2-fold rotation along the $x$ axis $\mathcal{C}_{2x}$, inversion $\mathcal{I}$, rotoinversion $\mathcal{S}_{6z}=\mathcal{C}_{3z} \mathcal{I}$ and mirror symmetry $\mathcal{M}_x=\mathcal{C}_{2x} \mathcal{I}$. At the (0001) surface of MnBi$_2$Te$_4$, since $\mathcal{C}_2$, $\mathcal{I}$ and $\mathcal{S}_6$ symmetries are broken, the point group is reduced to $\mathcal{C}_{3v}$. Therefore, when we take the local magnetic moment of Mn atoms into account, the symmetry preserved at this surface can only be $\mathcal{C}_{3z}$, $\mathcal{M}_x$, in-plane translation symmetry $\tau_{ab}$, $\mathcal{T}$ and their combinations, i.e., $\mathcal{C}_{3z} \mathcal{T}$, $\mathcal{M}_x \mathcal{T}$ and $\tau_{ab} \mathcal{T}$. We thus expect that the observed gapless surface Dirac cone is protected by one of these symmetries. Such a symmetry $\mathcal{G}$ should fulfill $\mathcal{G}^2 = -1$ in our spin 1/2 systems for at least certain high-symmetry points in the $k$-space, resulting in Kramers degeneracy and nontrivial topological classification. While high-order topology and gapless hinge/edge states can be protected by symmetry operations like 3-fold axis \cite{Schindler}, $\mathcal{M}_x \mathcal{T}$, $\mathcal{C}_{3z}$ and $\mathcal{C}_{3z} \mathcal{T}$ are excluded in our following analysis because they do not support gapless surface states \cite{ZhangLiu}. Hence, the remaining symmetry operations are $\mathcal{T}$, $\mathcal{M}_x$ and $\tau_{ab} \mathcal{T}$.

\begin{table*}[t]
\caption{Properties of MnBi$_2$Te$_4$ with different magnetic configurations considered in our work, including the calculated total energy with respect to the magnetic structure of ground state, i.e. A-type AFM ($z$) configuration, the gap size of surface states, the topological classification and the corresponding symmetry. The number in the parenthesis is the square of the symmetry operation.}
\centering
\begin{tabular}{ccccc}\hline\hline
Phase & Energy & Surface $E_g$ & Topological & Symmetry \\
  & (meV/Mn) & (meV) & classification & \\ \hline
A-type AFM ($z$) & 0.00 & 62 & $\mathbb{Z}_2$ & $\mathcal{S}^c$ $\left(-e^{-2ik\tau_{1/2}^c}\right)$ \\	
A-type AFM ($y$) & 0.41 & 0.3 & - & - \\
A-type AFM ($x$) & 0.41 & 0 & MCN & $\mathcal{M}_x$ $(-1)$ \\
G-type AFM & 8.34 & 0 & $\mathbb{Z}_2$ & $\mathcal{S}^a$ $\left(-e^{-2ik\tau_{1/2}^a}\right)$ \\
C-type AFM & 8.38 & 0 & $\mathbb{Z}_2$ & $\mathcal{S}^a$ $\left(-e^{-2ik\tau_{1/2}^a}\right)$ \\
PM & 5.73 & 0 & $\mathbb{Z}_2$ & $\mathcal{T}$ $(-1)$ \\
NM & $4.12 \times 10^{3}$ & 0 & $\mathbb{Z}_2$ & $\mathcal{T}$ $(-1)$ \\   \hline \hline
\label{Table1}
\end{tabular}
\end{table*}

The first possibility is that the ordered spin in the bulk state may show fragility at the surface layers, leading to a lower transition temperature and a PM state with spin disorder. In this case the electron hopping and magnetic moment decouple with each other and $\mathcal{T}$ symmetry restores, leading to gapless surface Dirac cone from a 3D $\mathcal{T}$-preserved TI. To obtain the total energy of such a PM state, we use a $4 \times 4$ supercell with special quasirandom structures (SQS) to simulate the spin disorder effect \cite{Zunger}. Such approach provides the best choice of randomness for a finite-size supercell in guaranteeing the best match of correlation functions of the infinite alloy, and thus can be applied in both cases of composition disorder and spin disorder. The total energy of such magnetic configuration is found to be 5.7 meV/Mn compared with the ground state.

Without ordered magnetization, the model Hamiltonian for a single surface of MnBi$_2$Te$_4$ can be written up to the third order as $H=H_1+H_3$. The first term $H_1$ is the well-known surface Rashba term
\begin{equation} \label{Eqn1}
H_1=\alpha_1 \left(k_x \sigma_y-k_y \sigma_x \right),
\end{equation}
where $\boldsymbol{\sigma}$ is the Pauli matrix for spin degree of freedom. $H_3$ is the cubic-$k$ term written as
\begin{equation} \label{Eqn2}
H_3=\alpha_2 [ (k_x^2+k_y^2 ) k_x \sigma_y-(k_y^2+k_x^2 ) k_y \sigma_x ]+\alpha_3 (k_x^2-3k_y^2 ) k_x \sigma_z,
\end{equation}
which is derived from the $k \cdot p$ perturbation by considering all the possible $k$-polynomial terms that respect the crystal symmetry \cite{FuLiang}. The basis used for this $k \cdot p$ model is the $m_j=\pm1/2$ states belonging to the representation $E_{1/2}$ \cite{GroupTable}, because the bulk bands near fermi level are mainly contributed by Te-$p_z$ and Bi-$p_z$ orbitals with $l=1$ and $m_l=0$. The Hamiltonian $H$ apparently support a gapless Dirac cone at the $\bar{\Gamma}$ point. Fig. \ref{Fig4}(b) shows the gapless surface Dirac cone of nonmagnetic MnBi$_2$Te$_4$ calculated by DFT as an approximation of the PM state, indicating a 3D $\mathcal{T}$-preserved TI. A careful comparison between ARPES data and DFT calculation reveals good consistency for the spin disorder scenario \cite{Supplement}, which also explains the observed almost unchanged band structure with the temperature across the transition point of 24 K, and the robustness of the surface Dirac cone against severe degradation.

The second possible type of symmetry that can protect the gapless surface state is the mirror symmetry. In this case, we begin with A-type AFM with in-plane magnetic moment that is perpendicular to one mirror plane $\mathcal{M}_x$ (because of $\mathcal{C}_{3z}$, there are three equivalent mirror planes at the (0001) surface). Such configuration only needs to overcome a small magnetic anisotropy energy compared with the magnetic ground state, i.e., 0.4 meV/Mn. In this case, MnBi$_2$Te$_4$ is calculated to be an AFM TI and an $\mathcal{M}_x$-protected AFM topological crystalline insulator (TCI) with nontrivial mirror Chern number (MCN) - in other words, a dual AFM TI. This is analogous to Bi$_2$Se$_3$ as a nonmagnetic dual TI \cite{Rauch}. As a result, the TCI phase gives rise to a gapless Dirac cone slightly off the $\bar{\Gamma}$ point, as shown in Fig. \ref{Fig4}(c). In MnBi$_2$Te$_4$ the shift is 0.005 $\textrm{\AA}^{-1}$ along the $k_y$ direction that is perpendicular to magnetic moment. Note that if this is realized at the surface, and the sizes of the magnetic domains having opposite magnetic moments are smaller than the size of the ARPES incident beam ($\sim5$ $\mu$m for laser ARPES), we would expect a doubling of the Dirac cone surface states with momentum separation ~0.01 $\textrm{\AA}^{-1}$, which is marginally detectable under the momentum resolution of our ARPES data. The fact that this is not observed in our dataset would indicate either the absence of in-plane A-type AFM at the surface, or magnetic domains that are significantly larger than $\sim5$ $\mu$m.

We next consider the magnetic moment along a more general in-plane direction. Without the protection of mirror symmetry, we can expect gap-opening at the (0001) surface. This can be understood by adding a magnetization-induced Zeeman term $H_{mag} = g \mathbf{B} \cdot \boldsymbol{\sigma}$ with in-plane magnetic field to the nonmagnetic Hamiltonian $H$ (see Eqs. \ref{Eqn1}, \ref{Eqn2}). If $H_3$ is omitted, $H_1+H_{mag}$ will lead to the shift of the Dirac point perpendicular to the direction of magnetic moment without gap opening. On the other hand, adding $H_{mag}$ to $H = H_1  + H_3$ will open a gap at the Dirac cone because the inclusion of the high-order term $H_3$ introduces a hexagonal warping term $\sigma_z$ that tilts the spin out of the plane \cite{FuLiang, Rauch}. Only if $\mathbf{B}$ is perpendicular to one of the mirror plane ($k_x=0$ or $k_x = \pm \sqrt{3} k_y$), the $\sigma_z$ term vanishes at the shifted Dirac point and the gapless Dirac cone retains. As shown in Table \ref{Table1}, the features of the surface band gaps with different in-plane magnetic moment derived from the model Hamiltonian is consistent with our DFT calculation. However, our DFT calculation reveals that the size of the gap induced by high-order term $H_3$ is as small as $\sim0.3$ meV, which is negligible within the resolution of ARPES. Hence, a general A-type AFM configuration with in-plane magnetic moment is also compatible with our experimental observation. Since the total energies for A-type AFM with different in-plane spin orientations are almost the same, we suggest that the surface layers of real samples probably consist of magnetic domains with different in-plane FM spin moments.

The third type of symmetry is the combination of $\mathcal{T}$ and translation symmetry $\tau_{ab}$. This type of symmetry requires intralayer AFM spin configuration, exemplified by the G-type AFM where both intralayer and interlayer alignment of magnetic moments are AFM, as shown in Fig. \ref{Fig4}(d). In such a magnetic configuration there exist $\mathcal{S}^i = \mathcal{T} \tau_{1/2}^i$ symmetries along all the three lattice vectors $i = a, b, c$ that correspond to two independent elements $\mathcal{S}^a$ and $\mathcal{S}^c$ in the magnetic space group ($\mathcal{S}^a$ and $\mathcal{S}^b$ are equivalent elements). The square of $S^i$ equals to -1 at certain time-reversal invariant momenta that meet the requirement $k \cdot \tau_{1/2}^i = n\pi$, leading to two $\mathbb{Z}_2$ invariants in this system. Since the only band inversion occurs at the $\Gamma$ point in G-type AFM MnBi$_2$Te$_4$, all the three topological invariants are nontrivial, giving rise to gapless Dirac cone at the (0001) surface because $\mathcal{S}^a$ remains invariant at this surface. Similarly, some other magnetic configurations with in-plane AFM alignment, such as C-type AFM, also host gapless (0001) surface state protected by $\mathcal{S}^a$ symmetry with in-plane fractional translation. However, our DFT calculations shows that forming intralayer AFM alignment in the surface layers need to overcome the favored intralayer FM exchange coupling in the bulk, leading to a large energy cost of about 8.3 meV/Mn \cite{Supplement}. Furthermore, if the intralayer AFM is realized at the surface, the in-plane Fermi surface should exhibit band folding effect, which is not observed by our ARPES measurements. Overall, to obtain the direct evidence for addressing the origin of the gapless feature, it is crucial to involve surface-sensitive detection, such as X-ray magnetic circular dichroism and the NV center technique, to determine the surface magnetic structure of such intrinsic magnetic TI in the future.

\subsection{V. Discussion and Conclusion}

Apart from the possible origins of the gapless surface Dirac cone due to surface spin reorientation in ideal MnBi$_2$Te$_4$ single crystals, we briefly discuss the possibility of structural deformation owing to the sample imperfection that could also hint the deviation of the electronic structure between ARPES measurements and the theoretical calculations based on A-type AFM magnetic configuration. For example, the MnBi$_2$Te$_4$ thin-film could be grown by co-evaporating Mn and Te elements onto Bi$_2$Te$_3$ surface, corresponding to the coverage of a MnTe layer \cite{Gong_CPL}. Although the Mn$^{2+}$ ion is energetically favorable to intercalate into the central layer of a MnBi$_2$Te$_4$ SL, the growing condition determines the existence of cation mixing between Mn and Bi, or antisite defects \cite{Zeugner}. Such occasion would lead to disordered Mn distribution when approaching the surface layers as well as magnetic disorder. In addition, it is reported that the synthetic MnBi$_2$Te$_4$ tends to experience a decomposition into Bi$_2$Te$_3$ and MnTe$_2$ phases at a higher temperature \cite{LeeDS}. Taking into account the surface potential, the surface MnBi$_2$Te$_4$ layer might also suffer structural reconstruction or dislocation, such as decaying to a Bi$_2$Te$_3$ layer, rendering different band dispersions compared with ideal MnBi$_2$Te$_4$ surface. At last, the imperfect crystal could lead to a number of ferromagnetic domains of which the moments point to different directions, contributing different values of the Zeeman term $H_{mag} = g \mathbf{B} \cdot \boldsymbol{\sigma}$. As a result, the measurable topological surface states could be compensated by the synergetic effect of different domains.

In summary, we demonstrate unambiguously by systematic ARPES measurements that MnBi$_2$Te$_4$ is not an ideal AFM TI. A gapless surface Dirac cone exists experimentally in single crystal MnBi$_2$Te$_4$, hosting massless Dirac quasiparticles. The Dirac cone is found to be quasi-2D and stable under massive surface absorbents; the bulk and surface electronic structures show no observable change across the bulk magnetic transition temperature. Our first principles calculation further identified several magnetic configurations that could yield such a gapless topological state. It is important to note that such reconstruction might not be limited only in MnBi$_2$Te$_4$, but in a group of other intrinsic magnetic topological materials, because several competing phases with different structural/magnetic orders could have small energy differences and thus are sensitive to surrounding environments and the presence of surface termination. Therefore, our research also sheds light on other magnetic topological materials, revealing that nature deals with AFM TIs in a more intricate way than previously thought. Motivated by our results, the theme of future works should be finding a way to overcome such surface reconstruction in favor of the long-sought axion insulators, or make use of such reconstruction to build devices with novel transport phenomena. Taking collectively, our discovery of the unexpected massless Dirac quasiparticles in an AFM TI indicates a space-varying magnetic structure more complex than previously assigned in these materials, brings about a more complete description of magnetic topological systems, and paves the way for the exploration of the interplay of massive and massless Dirac particles in a single material platform.

\begin{acknowledgments}
We thank Jieming Sheng, Yue Zhao, Haizhou Lu, Jianpeng Liu, Weiqiang Chen and Dapeng Yu for helpful discussions. ARPES experiments were performed with the approval of the Hiroshima Synchrotron Radiation Center (HSRC), Hiroshima, Japan under Proposal Nos. 19AG014 and 19AG004. Work at SUSTech was supported by the National Natural Science Foundation of China (NSFC) (Nos. 11504159 and 11874195), NSFC Guangdong (No. 2016A030313650), the Guangdong Innovative and Entrepreneurial Research Team Program (Nos. 2016ZT06D348 and 2017ZT07C062), the Shenzhen Peacock Plan Team (No. KQTD2016022619565991), the Shenzhen Key Laboratory (No. ZDSYS20170303165926217), and the Technology and Innovation Commission of Shenzhen Municipality (Nos. JCYJ20150630145302240 and KYTDPT20181011104202253). Work at UCLA was supported by the U.S. Department of Energy (DOE), Office of Science, Office of Basic Energy Sciences under Award Number DE-SC0011978.

Y.-J. H., P. L. and Y. F. contributed equally to this work.
\end{acknowledgments}

\end{document}